\documentstyle[aas2pp4,flushrt]{article}

\tighten
\lefthead{Giallongo et al.}
\righthead{The Photometric Redshift Distribution}

\begin{document}

\title{
THE PHOTOMETRIC REDSHIFT DISTRIBUTION AND EVOLUTIONARY PROPERTIES 
OF GALAXIES UP TO $z\sim 4.5$ IN THE FIELD OF THE QUASAR BR1202-0725
\footnote{Based on observations collected at the NTT 3.5m of ESO and
at UH 2.2m of the Hawaii}}

\bigskip
\bigskip

\author{E. Giallongo$^1$, S. D'Odorico$^2$,
A. Fontana$^1$, S. Cristiani$^3$,\\
E. Egami$^4$, E. Hu$^5$,  R. G. McMahon$^6$,
}

\bigskip
\bigskip
\bigskip
\bigskip

\noindent
$^1$ Osservatorio Astronomico di Roma, via dell'Osservatorio, I-00040
Monteporzio, Italy\\
\noindent
$^2$ European Southern Observatory, Karl Schwarzschild Strasse 2,
D-85748 Garching, Germany\\
\noindent
$^3$ Dipartimento di Astronomia, Universit\`a di Padova, vicolo
dell'Osservatorio 5, I-35122, Padova, Italy\\
\noindent
$^4$ MPE, Postfach 1603, D-85740 Garching, Germany\\
\noindent
$^5$ Institute for Astronomy, University of Hawaii, 2680 Woodlawn Dr.,
Honolulu, HI 96822, USA\\
\noindent
$^6$ Institute of Astronomy, Cambridge CB3 0HA, UK\\

\newpage
\begin{abstract}

We present a deep BVrIK multicolor catalog of galaxies in the field of
the high redshift ($z=4.7$) quasar BR 1202-0725. Reliable colors have
been measured for galaxies selected down to $R=25$. Taking advantage
of the wide spectral coverage of the galaxies in the field, we compare
the observed colors with those predicted by spectral synthesis models
including UV absorption by the intergalactic medium and dust
reddening. The choice of the optical filters has been optimized to
define a robust multicolor selection of galaxies at $3.8\leq z\leq
4.5$.  Within this interval the surface density of galaxy candidates
with $z\sim 4$ in this field is 1 arcmin$^{-2}$.  Photometric
redshifts have been derived for the galaxies in the field with the
maximum likelihood analysis using the GISSEL library of $\sim 10^6$
synthetic spectra. The accuracy of the method used has been discussed
and tested using galaxies in the Hubble Deep Field (HDF) with known
spectroscopic redshifts and accurate photometry.  A peak in the
redshift distribution is present at $z\simeq 0.6$ with relatively few
galaxies at $z>1.5$. At variance with brighter surveys ($I<22.5$)
there is a tail in the distribution towards high redshifts up to
$z\sim 4$. The luminosity function at $z\sim 0.6$ shows a steepening
for $M_B>-19$.  This increase is reminiscent of that found in the most
recent estimates of the local luminosity function where a similar
volume density is reached about 2 mag fainter. The observed
cosmological ultraviolet luminosity density is computed in the overall
redshift interval $z=0.3-4.5$ reaching a value $\sim 2\times 10^{19}$
W Hz$^{-1}$ Mpc$^{-3}$ at $z\sim 0.8$.  Including recent local
estimates it appears that the UV luminosity density changes by a
factor $\sim 2.5$ in the overall redshift interval $z=0.1-4$, not
including correction for fainter undetected galaxies. Thus we find
that the evidence of a marked maximum in the luminosity density at
$z\sim 1-1.5$ for galaxies with $R\leq 25$ is weak.  We have derived
in a homogeneous way, using the GISSEL libraries, the physical
parameters connected with the fitted spectral energy
distributions. Thanks to this new approach, the problem of the star
formation history of the universe is dealt with in a self consistent
way taking into account the dust and metallicity distributions derived
from the spectrophotometric properties of each galaxy in the
sample. The bulk of the blue intermediate redshift population with
$z=0.4-1$ mostly consists of very young star-forming galaxies with a
median starburst age of the order of a few $10^8$ yr and typical mass
in luminous stars $\sim 2\times 10^8$ M$_{\odot}$. The presence of
this young population is in contrast with pure luminosity evolutionary
model (PLE) based on a single high formation redshift.  The
cosmological mass in formed stars per unit comoving volume at $z\sim
3$ is already $\sim 20$\% of that formed at $z=0.5$ in our magnitude
limited sample. Predictions based on the standard hierarchical
clustering models are smaller, although not far from that derived from
the observations.

\end{abstract}

\bigskip

\twocolumn

\section{INTRODUCTION}

Deep photometric and spectroscopic surveys of galaxies performed by
means of large ground-based telescopes and the HST are providing a
first insight into the cosmological distribution of galaxies of
different spectral types up to $z\sim 5$.

The bulk of the contribution to surveys of intermediate depth (B$<24.5$,
$I<23$, and $K<20$) comes from galaxies at redshifts $z\sim 0.5$ with
a tail in the distribution extending up to $z\sim 1.5$ (Songaila et
al. 1994, Lilly et al. 1995, Ellis et al. 1996, Cowie et
al. 1996). These surveys show substantial evolution in the galaxy
properties starting already at $z\geq 0.5$.  The presence of
massive star formation is provided by the observed blue colors of
galaxies and by the detection of strong emission lines.

The recognition of the evolutionary path of the average star formation
rate per unit comoving volume is one of the main tools to understand
the processes that control the formation and evolution of galaxies in
the universe. An increase by a factor of ten in the redshift interval
$z=0-1$ has been recently suggested (Lilly et al. 1996, Madau et
al. 1996). This of course implies that about half of the stars formed
at intermediate redshifts, in general agreement with theoretical
expectations of CDM cosmologies. Nevertheless, the same cosmological
scenarios predict a fraction $<2$\% of the present mass density in
stars at $z>4$ (Cole et al. 1994). It is therefore at these high
redshifts that cosmological scenarios for galaxy formation are more
vulnerable to observational constraints.  Thus, the search and
discovery of galaxies at $z>2$ is a major challenge of recent
observational cosmology.

Until recently, however, these objects have proven to be quite
elusive, beeing detected neither in complete spectroscopic surveys
limited to $I\leq 22.5$, (Lilly et al 1995 CFRS, Cowie et al 1996),
nor in field surveys based on the search of highly redshifted
Ly$\alpha$ line emission (Thompson et al 1995) with some relevant
exceptions (Macchetto et al. 1993, Warren and M$\o$ller 1996).

Quite recently, improved sensitivities and new observational
strategies have led to the discovery of star-forming galaxies up to
$z\sim 4.5$. Deep Ly$\alpha$ searches have been successfully tuned at
particular redshifts, such as those of previously known quasars (Hu
and McMahon 1996), radio galaxies (Pascarelle et al 1996) or strong
QSO absorption systems (Francis et al 1996).  Deep multicolor searches
are also proving very effective at detecting field galaxies in
relatively wide redshift intervals. The technique based on the color
selection of the Lyman break feature in flat spectrum star-forming
galaxies has been applied by Steidel and co-workers to look for field
galaxies at $2.8 < z < 3.4$, (Steidel and Hamilton 1992, Steidel et al
1995), with a remarkable spectroscopic success rate of $\simeq 70$\%
(Steidel et al 1996). We have also applied a similar multicolor
technique to look for the galaxies responsible for strong Ly$\alpha$
absorption in quasar spectra. By means of deep BVrI images around the
$z=4.695$ quasar BR1202-0725 we discovered a galaxy companion whose
redshift was estimated to be in the range $z\geq 4.4-4.7$ (Fontana et
al. 1996) on the basis of sharp color criteria obtained comparing
observed colors with that expected by star-forming galaxies once UV
absorption by the intergalactic medium has been taken into account
(Madau 1995). Narrow band imaging and spectroscopic follow up have
confirmed the $z=4.7$ redshift of the galaxy (Hu et al. 1996,
Petitjean et al. 1996).

The majority of galaxies detected in the Hubble Deep Field (HDF;
Williams et al. 1996) are too faint to be investigated by
spectroscopy.  The only way to derive statistical information about
their cosmological distribution is based on color-estimated redshifts
(Gwinn \& Hartwik 1996, Lanzetta et al. 1996, Sawicki et al. 1997).
The reliability of the method depends mainly on the number of colors
used and on the modeling of the galaxy templates. The possibility to
calibrate the color-estimated redshifts with fair samples of
spectroscopic galaxies allows better accuracy (Koo 1985, Koo \& Kron
1992, Connolly et al. 1995).

We have applied the ``photometric redshift technique'' to derive the
redshift distribution of the complete galaxy sample in the field of
the quasar BR1202-072 and the results are discussed in this work.  To
make the redshift estimates more robust we have extended the spectral
coverage in the field adding deep K images ($K\simeq 21.5$ at
3$\sigma$ sky limit).  Photometric redshifts have been derived with a
maximum likelihood analysis over the five broad bands BVrIK using a
library of $\sim 10^6$ synthetic spectra. A calibration of the
method has been carried out over about 60 galaxies in the HDF with
known spectroscopic redshifts. In Sect. 2 we present the data sample,
in Sect. 3 the library of galaxy synthetic spectra and our statistical
technique for color-estimated redshifts. In Sect. 4 we describe our
multicolor selection of galaxies with $z>3.8$ and in Sect. 5,6 we
describe and discuss the relevant results concerning the cosmological
evolution of the galaxy properties.

\section{THE DATA SAMPLE}

The field of BR 1202-0725 was observed with the SUSI imaging CCD camera at
the Nasmyth focus of the ESO New Technology Telescope. Four broad band
filters were used during three photometric nights, 1995 April 23-36.
Standard BVI passbands of the Johnson-Kron-Cousins system and
Thuan-Gunn $r$ system were used. Several dithered images were obtained
in the four broad bands with total integration times of 15000, 10800, 7200,
7800 s for the BVrI bands. The FWHM of the stellar images in the combined
frames are 1, 1, 0.6, 0.65 respectively. This is better than most data
of other deep surveys and it is an advantage for the accuracy of the
photometry at the fainter magnitudes where the density of objects is high.

The photometric calibration was obtained from several standard stars
observed on the same nights and at similar airmasses. Total magnitudes
were obtained in an aperture of fixed size (1.8 arcsec), applying a
seeing-dependent aperture correction.

We have calibrated $B'V'R'I'$ magnitudes in the ``natural'' system
defined by our instrumental passbands. The zero points of our
instrumental system were adjusted to give the same $BVRI$ magnitudes
as in the standard system for stellar objects with $B-V=V-R=R-I=0$.
The colour transformation for stellar objects was $I'=I$,
$B'-V'=0.91\times (B-V)$, $V'-R'=1.17\times (V-R)$, and
$R'-I'=0.74\times (R-I)$. The magnitudes were corrected for galactic
absorption, with $E(B-V)=0.03$. Transformations to the AB systems
give $I_{AB}=I'+0.47$, $R_{AB}=R'+0.21$, $V_{AB}=V'$, $B_{AB}=B'-0.16$

The K band observations have been obtained at 2.2m telescope of the
University of Hawaii in two runs: 15-17 April 1994 for a total of 10.4
h on the central $90\times 90$ arcsec$^2$ field and 17-19 March 1995
with QUIRC 1024$^2$ array for a total of 7 hrs on a 3$\times 3$
arcmin$^2$ field encopassing the SUSI field.

A galaxy catalog has been extracted from the best quality image in the
$r$ band with the SExtractor image analysis package (Bertin \& Arnouts
1996).  To detect objects and compute total magnitudes we have
followed the standard procedure described by Smail et al. (1995). We
first smooth the frame with the point-spread function derived from
stellar profiles in the field and then threshold it at a level of 2.2
$\sigma$ of the sky noise in the raw frame and identify objects. This
threshold provides a detection of galaxies down to $R'=25.8$.

Total magnitudes have been obtained using a fixed aperture (1''.8
diameter or $\geq 3$ FWHM) after an aperture correction to a 5''
diameter under the assumption that the objects have roughly stellar
profiles, this results in an aperture correction of $\delta R'=0.1$ at
$R'\simeq 25$. The validity of the assumption has been tested in
previous similar deep images (Smail et al. 1995).

For objects brighter than $R'=23$, where isophotal diameters are
larger than our fixed aperture, we adopted isophotal
magnitudes. To measure colors for our objects we measure
aperture magnitudes in all the bands in the same positions defined in
the $r$ frame with the appropriate aperture corrections. To put
stringent limits on the galaxy colors we have adopted 2$\sigma$ upper
limits in the $B'V'I'$ bands of 27, 26.6, 25, respectively. Since the
sky noise in the $K$ image changes appreciably across the field, we
have adopted the more conservative 3$\sigma$ threshold $K=21.5$.

In order to discuss the statistical spectral properties of faint
galaxies we have computed reliable colors only for galaxies with
$R'\leq 25$ for which the surface density in our field 1.4$\times
10^5$ deg$^{-2}$, in agreement with the results of Smail et
al. (1995).

Coordinates (obtained using 11 APM stars as reference in the field)
and magnitudes for all the galaxies selected in the interval
$20<R'<25$ are shown in Tab. 1.

\section{THE PHOTOMETRIC REDSHIFT TECHNIQUE}
\subsection{The Synthetic Spectral Libraries}

To derive photometric redshifts for all the galaxies in the catalog we
have computed the expected colors (in our photometric system) for
galaxies as a function of the redshift using the spectral synthesis
model (GISSEL library) by Bruzual \& Charlot (1997). Models of this
kind have a number of free parameters which control the spectral
properties of different spectral types of present-day galaxies.

The UV flux emitted by galaxies is directly related to the
star-formation rate of each galaxy. This is governed by
the assumed $e-folding$ star formation time-scale $\tau$ and by the
age of the galaxy. Different combinations of ages and $\tau$ values
result in different spectral types. For example, a star-formation
timescale of $\tau \leq 1$ Gyr and an age $\geq 12 $Gyr is more
appropriate for an early type galaxy, while $\tau = 2, 4, 7, $ Gyr or
constant SFR fit well typical spectra of Sb, Sbc, Sc, or Irregulars
respectively (Bruzual \& Charlot 1993).

Although smooth star-formation rates declining with an exponential law can
adequately represent the global star-formation history of most of the
present day galaxies, there are observational hints and theoretical
expectations for a star-formation history which shows signs of
stochastic episode of starburst formation. 

This situation can be parameterized in terms of the Scalo (1986)
parameter $b$, defined as the ratio of the SFR at the current time to
the past SFR averaged over the age of the galaxy. A constant or
declining SFR of the kind $SFR\propto exp(-t/\tau)$ gives always
$b\leq 1$. In contrast, multiple bursts can lead to values of the
Scalo parameter $b>1$ as observed in a few local galaxies where
$b=1-5$ has been derived (e.g. Kennicutt, Tamblyn, \& Congdon 1994).

To represent in a simple way the galaxies with $b>1$ we have computed
a set of models consisting of two short bursts separated by a time
delay of 3 Gyr. The first burst has $\tau =0.3, 5$ Gyr, the second
burst $\tau = 0.3$ Gyr. In these clear examples most of the galaxy
mass in stars ($>70$\%) has been formed during the first burst but
almost all of the current star formation activity is due to the second
burst. In this way the UV flux is controlled by the current SFR while
the red and IR fluxes are largely controlled by the stellar population formed
in the past 3 Gyr during the first burst.  Time delays greater than 3
Gyr have not been considered since they are hard to infer from the
broad band spectral fits. Indeed, for values $>3$ Gyr the shape of the
red part of the galaxy spectrum does not change significantly.

This kind of star formation history can reproduce a current star
formation activity greater than the average past rate, with a Scalo
parameter in the range $b=1-4$.

We have allowed for models based on different shapes (Scalo, Miller-Scalo
or Salpeter) of the initial mass function (IMF) including stars in the
$0.1 < M < 65$ M$_{\odot}$ range. We have also considered the case of
metallicity $Z= Z_{\odot}$, $Z=0.2 Z_{\odot}$ and as low as $Z=0.02
Z_{\odot}$ since recent results on the spectral properties of the
brighter $z\sim 0.5-1$ galaxies in the CFRS sample suggest the
presence of non solar metallicities in a fraction $\sim 30$\% (Hammer
et al. 1997).

Two important features that alter the UV flux of high redshift
galaxies must be added to the spectral synthesis models: the
reddening produced by internal dust and Lyman series absorption
produced by the neutral fraction of the intergalactic medium (IGM)
intervening along the line-of-sight to each galaxy. While the former
effect can be already appreciable at intemediate redshifts, the latter
becomes important in the observed U,B,V bands only at $z>2,2.6,3.5$
respectively.

We have adopted the attenuation law by Calzetti et al. (1994,1997)
which is derived from the observed local starburst galaxies and seems
more appropriate when the absorption by the interstellar medium within
each galaxy is clumpy, as expected in high $z$ galaxies. This
reddening curve is more grey than the Milky Way curve and does not
show any 2000 {\AA} bump.

For the Lyman series absorption produced by the IGM we have
adopted the average HI opacity of a clumpy universe as a function of
redshift computed by Madau (1995) where the average attenuation
e.g. in the B band is $\sim 1$ mag at $z\sim 3.5$ and $\sim 2$ mag at
$z\sim 4$.

Finally, at any redshift  galaxies are allowed to have any age smaller
than the Hubble time at that redshift ($\Omega =1$ and $H_0 = 50$ km
s$^{-1}$Mpc$^{-1}$ have been adopted throughout the paper).  A large
dataset ($\simeq 10^6$) of ``simulated galaxies'' has thus been
produced and compared with the observed colors. The full grid in the
parameter space is shown in Tab. 2.

To show the main differences in the new GISSEL library due to the
different IMF and metallicities assumed, we have plotted in Fig.~1
various colors as a function of redshifts for a galaxy forming stars
at a constant rate from $z=5$ with metallicities $Z=0.2,1 Z_{\odot}$
and assuming the Scalo or Salpeter IMFs.  We note that the colors
investigated here depend on the IMF rather than on metallicity for
$0.1\leq Z/Z_{\odot}\leq 1$. In particular, there are no appreciable
changes in the $(R-I)_{AB}$ and $(V-I)_{AB}$ colors of $z>2$ galaxies
varying both metallicities and the IMF since the spectral shape for
$\lambda <4000$ {\AA} rest-frame is similar at different
metallicities. Differences in metallicity produce appreciable changes
only in the $I_{AB}-K$ colors at $z<1.5$.  In contrast, $I_{AB}-K$ and
$(B-I)_{AB}$ colors at a given $z<1.5$ appear redder in the Scalo
case.

The allowance for the presence of dust in the fitting models has some
consequence on the estimated redshift distribution and on the
cosmological evolution of the derived physical properties like the
average SFR.  In fact the presence of dust does not change appreciably
the redshift estimate for $z<1.4$ or $z>2.8$ where we have strong
color gradients due to the presence of the 4000 {\AA} break or the
Lyman absorption respectively.  Nevertheless there is a specific
redshift interval at $1.5\leq z \leq 2.8$ where the expected dust free
spectrum of any star forming galaxy is flat everywhere from the K down
to the B band.  Since the B band samples the 1500 {\AA} rest frame
region at $z=2$, even a modest amount of dust ($E_{B-V}=0.1$ in the
rest frame) produces an appreciable reddening which shifts the
dust-free redshift estimate from the true $z\sim 2$ toward the
$z\leq 1.4$ region. Thus the inclusion of dust in the model computations
reduces systematic errors present in dust-free estimates of the
photometric redshifts.

A general conclusion that can be drawn is that the selection of high
$z$ candidates ($z>2.5$) does not depend strongly on the particular
IMF or metallicity adopted, as discussed in Sect. 4. At $z<1.5$ on
the contrary, the photometry can provide information on the physical
properties of the model, once the spectroscopic redshifts are known.

\subsection{Statistical analysis}

To obtain statistical information on the photometric redshifts of all
the galaxies in the catalog we have applied a best-fitting procedure
which is able to give the most appropiate spectral template matching
the observed galaxy colors.

This ``photometric redshift'' approach is known to work successfully
for low redshift bright galaxy samples where extensive calibration
with spectroscopic redshifts is available (Sawicki et al. 1997,
Lanzetta et al. 1997; see also Connolly et al 1995).

The classical $\chi ^2$ fitting procedure allows to take into account
for redshift uncertainties due to photometric errors and to
the similarity of model spectra at very different redshifts.

For each template we computed
\begin{equation}
\chi ^2 = \sum_i \left[ {F_{observed,i}-s\cdot F_{template,i} \over \sigma _i}
\right]^2
\end{equation}
where $F_{observed,i}$ is the flux observed in a given filter $i$,
$\sigma _i$ is its uncertainty, $F_{template,i}$ is the flux of the
template in the same filter and the sum is over the 5
filters used in the field. The template fluxes have been normalized
to the observed ones with the factor $s$.

When a given object is not detected in a particular filter because it
is too faint we adopted the simple and conservative approach where all
the models with theoretical fluxes below the flux limit are accepted,
while the models with theoretical fluxes above the flux limit are
weighted assuming that the flux observed is the flux limit. The
best-fit redshift for each galaxy is reported in Tab. 1 together with
its reduced $\chi_{\nu}^2$ value. The redshift uncertainties
($z_{max}-z_{min}$ showing the 68\% confidence interval) are obtained
adopting the standard $\Delta \chi^2=1$ increment for the single
parameter being estimated.  In using the $\Delta \chi^2$ increment, a
minimization with respect to the other parameters has been performed
(Avni 1976). In this way, the resulting redshift uncertainties are
obtained taking into account for both the photometric errors and the
cosmic variance of the spectral templates.
 
The overall reliability of the method used is however given by the
comparison between photometric and spectroscopic redshifts as shown in
the next Section.

\subsection{Calibrating photometric redshifts with the HDF galaxies}

To assess the reliability of our technique we have compared
photometric and spectroscopic redshifts available for 55 galaxies in
the HDF for which infrared photometry is also available (Cowie 1997;
Cohen et al. 1996; Dickinson 1998; Lowenthal et al. 1997). Our
library of model spectra has been compared with this galaxy sample
using the five HST bands UBVIK for comparison with our five bands.
Since an accurate determination of galaxy colors is essential to
determine photometric redshifts with good accuracy, the galaxy
magnitudes in the HDF have been re-estimated using the Sextractor
package as in our field. Inspection of the HDF data shows that in
selected cases, the presence of substructures or close pairs with
different spectral characteristics is likely to affect the photometric
accuracy of the redshift estimates. This is particularly true if the
ground-based observations are taken in mediocre seeing conditions.

The calibration of the GISSEL models by means of spectroscopic
redshifts for the HDF galaxies allows a better definition of the
realistic space of the parameters involved. For a given IMF, the main
parameters which should be controlled to avoid unrealistic model
spectra are the dust reddening, age, metallicity and star-formation
time-scale. Peculiar combinations of these parameters can introduce
significant errors in the redshift estimates.  This would be the case
e.g. of models with a metallicity much less than the solar value but
an old age of the galaxy and a substantial dust reddening.  To be more
specific, a trend between metallicity and age has been introduced
where the case of extremely low metallicity ($Z=0.02 Z_{\odot}$) is
restricted to very young galaxies, in general agreement with both
observations and galaxy evolution models (e.g. Tantalo, Bressan \&
Chiosi 1997).

Moreover, a general correlation between metallicity and dust reddening
has been introduced in the sense that young galaxies with low
metallicity can be reddened only by a small amount of dust
($E_{B-V}\leq 0.05$).  This trend has been discussed by Calzetti et
al. (1994) as responsible for the correlation between the UV slope of
the power-law continuum and the galaxy metallicity. In their sample,
galaxies with low metallicities have a flatter UV spectrum on average,
as expected if the dust reddening is lower.

Adopting this recipe based on reasonable constraints supported by
observations of nearby galaxies we have obtained for the average
difference between spectroscopic and photometric redshifts in the HDF
sample $\sigma _z\sim 0.1$ in the redshift interval $z=0-3.5$
independently of the IMF. A value $\sigma _z\sim 0.2-0.3$ is obtained
when the unconstrained overall spectral library is used.

The relations between photometric and spectroscopic redshifts are
shown in Fig.~2 for all the IMF considered. As discussed above, an
average $\sigma _z=0.1 ~(0.15)$ is found for galaxies with $z<1.5
~(z<3.5)$ independently of the IMFs. A small trend in the average
$\langle \Delta z \rangle= \langle z_{spectr}-z_{phot}\rangle$ for
galaxies with $z<1.5$ has been observed depending on the IMF used. The
Salpeter IMF shows an average underestimate by $\langle \Delta
z\rangle=0.09$ which reduces to $\langle \Delta z\rangle =0.05$ for
the Miller-Scalo IMF and to $\langle \Delta z\rangle =-0.04$ for the Scalo
IMF.  Thus the maximum difference in the photometric redshifts
obtained by the different IMFs used is $\sim 0.1$ and an IMF shape
steeper than a Salpeter one seems favoured by the GISSEL spectral
models. In the next sections relevant results are given for the
Miller-Scalo IMF.

As discussed in Sect. 3.1 the inclusion of dust does not change in
general the discrimination between low ($z<2$) and high ($z>2$)
redshift galaxies. In particular a typical decrease in the redshift
estimate by $\Delta z\leq 0.2$ respect to the dust-free models is
obtained for the high $z$ HDF galaxies. The average amount of dust in
the high redshift HDF galaxies with $z>2$ depends of course on the IMF
used. The Salpeter IMF produces a larger fraction of massive stars
respect to the Scalo IMF and consequently a larger UV flux for a given
star formation rate. Thus, at $z\sim 3$ a larger amount of dust is
needed using the Salpeter IMF to reproduce the observed $U-K$ color.
Average values of $E_{B-V}=0.08, 0.11, 0.14$ are obtained for the Scalo,
Miller-Scalo and Salpeter IMFs respectively.

The inclusion of the $J,H$ bands allows a small improvement of the fit
in the region $0.8<z<1.5$ with $\sigma _z$ changing by a few
thousandths. However a so small improvement could be due in part to
the small number of galaxies with known spectroscopic redshifts in the
$z=1-2$ interval where the $J$ band samples the 4000 {\AA} break.

In conclusion, GISSEL spectral models can provide photometric
redshifts with comparable accuracy as obtained by means of the
observed templates collected by Coleman, Wu, \& Weedman (1980).
Although different empirical approachs to derive photometric redshifts
can provide better accuracy (Connolly et al. 1997), we have in this
case the advantage to explore the spectral and physical properties of
the galaxies derived from the GISSEL models.

This result is essentially due to the improvement of the new GISSEL
library in the UV part of the model spectra (Bruzual \& Charlot 1997)
and to the inclusion of internal reddening by interstellar dust in
galaxies and by the intergalactic neutral hydrogen absorption present
along the line of sight.

\section{GALAXIES AT VERY HIGH REDSHIFTS}

As a first step we use the predicted spectral properties of the high
redshift, star-forming galaxies to define a robust color selection
for candidates in the redshift interval $3.8<z<4.4$. The bright
fraction of these candidates represents an important sample for
spectroscopic investigation. We then compare this method with
the more sophisticated photometric redshift technique to check the
robustness of the multicolor selection.

A very efficient method to select galaxies which are actively forming
stars at high redshifts has been proposed by Steidel \& Hamilton
(1992). It is based on the detection of the Lyman break present in the
flat rest-frame UV continuum of these galaxies.  Steidel and Hamilton
(1992) used a particular set of U G R filters to select the Lyman
break galaxies in the redshift interval $2.8<z<3.4$.  Since the
spectrum of an actively star--forming dust--free galaxy is flat
longward of the Lyman break, an average color of G--R$\sim 0.5$ is
expected in the selected redshift interval. At the same time strong
reddening is expected in the U--G color which samples the drop of the
emission shortward of the Lyman break (U--G$>1.5$) for $z\sim 3$
galaxies.

Steidel et al. (1996) confirmed with low resolution spectra at the
Keck telescope the identification of the candidates in the expected
redshift interval proving the high success rate ($>70$\%) of this
multicolor selection. Extrapolating the success rate obtained for the
subsample of their candidates, they provided a first estimate of the
surface density of galaxies at $z\sim 3$ of the order of 0.4
arcmin$^{-2}$.

Their $U G R$ photometric system however limits the maximum redshift
for the Lyman break galaxies to $z\simeq 3.5$. Indeed at higher
redshifts the Lyman break enters the G band causing a reddening of the
$G-R$ colors and an increasing contamination by low redshift galaxies.

\subsection{The multicolor selection of the $3.8<z<4.5$ galaxies}

To select galaxies at $z\sim 4$ it is better to sample the complex shape
of the universal opacity to the UV photons of the intergalactic
medium. We have plotted in Fig.~3 the spectrum of a dust-free galaxy
forming stars at a constant rate at $z=3.25$ or $z=4.25$ (Bruzual \&
Charlot 1997), including the average attenuation due to the IGM
absorption (Madau 1995, Madau et al. 1996).  It is clear that the most
prominent spectral features come from the depression due to the
average Lyman series absorption and from the cutoff at the continuum
Lyman edge.  While the Ly$\alpha$ forest absorption produces a
fractional decrement of only $\sim 30$\% at $z\simeq 3$, at $z\sim
4.5$ about 60--70\% of the galaxy emission is lost causing a strong
and easily detectable reddening at the blue frequencies.

An efficient sampling of this complex absorption requires at least
four broad band filters in the optical range.  We have chosen the Johnson
$BV$, Gunn $r$ and Kron--Cousin $I$ filters, that are shown as broken
lines in Fig.~3, superposed to the galaxy spectra.

In our natural AB photometric system described in Section 2 we
expect high redshift, dust-free galaxies to show a marked flat
spectrum through the $r,I$ bands dominated by star--forming emission
resulting in a $(R-I)_{AB}\simeq 0$ color. At the same time we expect
an increasing reddening of the spectrum below 5500 {\AA} for a $z\sim
4$ galaxy because of the IGM absorption. The corresponding colors will
be large in $(V-I)_{AB}$ because of Ly$\alpha$ absorption and even
larger in $(B-I)_{AB}$ colors when the flux in the $B$ band is cut by
Lyman continuum absorption (Fig.~3).

Thus high redshift galaxies appear well segregated in our color space
(Fig.~1) independently of plausible ranges of the model parameters
involved.  The $(R-I)_{AB}$ color is sampling the intrinsic spectrum
of galaxies in a wide redshift interval up to $z=4.4$, where the IGM
absorption in the $r$ band begins to produce an appreciable reddening
in the $(R-I)_{AB}$ color.  Thus, in the redshift range $1.5<z<4.4$
all star--forming galaxies show $(R-I)_{AB}\sim 0\pm 0.1$.  On the
other hand, IGM absorption produces strong reddening first in
$(B-I)_{AB}$~ ($(B-I)_{AB}>$1.4 at $z> 3.5$), then in $(V-I)_{AB}$~
($(V-I)_{AB}>$0.7 at $z> 4$). Thus, the simultaneous presence of the
three colors at the average expected values can select star forming
galaxies with small dust reddening at $z \geq 4$ (Fontana et
al. 1996, Giallongo et al. 1996).

At $z<1.5$ the $(R-I)_{AB}$ colors are sampling progressively longer
rest-frame wavelengths, where the galaxy spectra are in general
steeper, always resulting in $(R-I)_{AB}>$0.2.  Any possible
contamination by an old population with steep blue spectra (a
pronounced 4000 {\AA} break) producing red $(B-I)_{AB}$ and
$(V-I)_{AB}$ colors at $z=0.5-1$ can be avoided just requiring a
``flat'' $(R-I)_{AB}$. The availability of deep $K$ band photometry in the
field helps to constrain even more the colors of $z\sim 4$ galaxies
which would always have $I_{AB}-K<2.5$ colors.

Taken all together, these results may be used to define relatively
sharp color criteria that select high redshift galaxy candidates,
similar to what was done by Steidel et al 1995 and Madau et al
1996. In our case, star forming galaxies in the redshift interval
$3.8\leq z \leq 4.4$ can be selected by imposing $(R-I)_{AB}\leq 0.1$,
$(V-I)_{AB}\geq 0.5$, $(B-I)_{AB}\geq 2$.  By adding the constraint on
the $I_{AB}-K<2$ color and allowing for $(R-I)_{AB}>0.2$, galaxies at
$z> 4.5$ can be selected.

Following this multicolor selection criterion we have identified 6
objects as star--forming galaxy candidates at $3.8< z\leq 4.4$, taking
into account upper limits in the B band. The spectroscopic follow--up
of these objects is difficult even for 10m-class telescopes. At the
typical magnitude and redshifts of our objects, the UV absorption
lines fall in the spectral region affected by strong sky emission and
can hardly be detected.

We stress here that even small amounts of dust can produce a
systematic overestimate of the redshift for the $z>3$ galaxies. As a
result, a dust-reddened galaxy with $E_{B-V}\simeq 0.1$ at $z\sim 3$
can mimic typical colors of a dust-free galaxy at $z\sim 3.5$. For
this reason it is difficult to evaluate from a selection in the
color-color plane, the surface densities of the $z\sim 3$ and $z\sim
4$ galaxies, separately, and a more sophisticated best fitting
procedure involving observed and predicted colors including dust
reddening should be pursued as described in the next Section.

\subsection {The photometric redshift distribution of\\ galaxies up to 
$z\sim 4.5$}

The photometric redshift distribution of the galaxies in our field is
shown in Fig.~4.  A peak in the redshift distribution is present at
$z\sim 0.6$ with few galaxies at $z>1.5$. This implies that at
the magnitude limit for spectroscopic follow up ($R_{AB}\sim 25.5$),
the bulk of the faint galaxies is at redshift $z<1$ and any
information about the luminosity function of galaxies at $z\geq 2$ can
be obtained by means of large and deep surveys of galaxies having
reliable photometric redshifts.

However, at variance with brighter surveys limited at $I<22.5$ a tail
in the $z$ distribution appears for $z>1.5$ and extending to $z\sim
4.5$. The broad band energy distributions of some of them are
displayed in Fig.~5, together with the best--fitting spectrum. As it
can be recognized, the high~$z$ identification is provided by the
large $(B-I)_{AB}$ and $(V-I)_{AB}$ colors sampling the IGM absorption
provided that $(R-I)_{AB}$ and $I_{AB}-K$ colors are consistent with
the flat spectrum of a star-forming galaxy.

We find 8 galaxy candidates in our field at $2.8\leq z\leq 3.5$
corresponding to a surface density of 1.6 arcmin$^{-2}$. This is 4
times larger than the recent estimates of Steidel et al. (1996) in the
same redshift interval. Given the small area of our field (4.84
arcmin$^2$) fluctuations in the surface density of high redshift
galaxies can explain this difference.  Indeed our estimate is in good
agreement with the number of spectroscopically confirmed galaxies
found by Lowenthal et al. (1997) in the HDF at a similar depth
$I_{AB}\leq 25.3$. They found 7 confirmed galaxies in the HDF. Thus
the surface density of $z\sim 3$ galaxies with $R'\leq 25$ (or
$I_{AB}\leq 25.3$) is about 1.6 arcmin$^{-2}$ both in our field and in
the HDF.

The derived surface density at $3.5< z\leq 4.4$ in our field is
lower. Indeed we find 5 of the 6 objects selected by our multicolor
selection (see Sect. 4.1), one object being lost because was at the edge
of our color selection. The surface density in this redshift interval
is 1 arcmin$^{-2}$.  In these estimates we have excluded the $z\sim 4.7$
galaxy companions discovered a few arcseconds away from the quasar
(Hu, McMahon, \& Egami 1996, Fontana et al. 1996, Petitjean et
al. 1996).

One further high redshift candidate (\#12 in our list) has been
excluded from our statistical analysis.  At the faint magnitude limit
of our sample, compact, late type stars have colors which can mimic
compact, high redshift galaxies. We have used a 150 m HST exposure of
this field obtained for a different program by three of us (R.McM.,
E.H. and E.E.) to study the morphology of our high redshift
candidates. Object \#12 is the only one with a stellar appearance. A
Keck spectrum obtained in the spring of 97 (Cowie and Hu, private
communication) does not rule out the identification with a late star.

Indeed, late type stars can show colors similar to that expected from
$z\sim 4$ galaxies. Moreover, the compact nature of the high redshift
galaxies (Steidel et al. 1996) prevents any a priori exclusion of the
high $z$ candidates based mainly on their morphological compacteness.
Therefore some small contamination could be present in the $z\sim 4$
candidates.  In our catalog however all the remaining $z>3.5$
candidates are clearly extended.

Finally, our estimate is a lower limit if a significant fraction of
$z>4$ star--forming galaxies have their intrinsic UV emission strongly
attenuated by dust extinction. Early type galaxies whose star
formation has happened at even higher redshifts in a short burst are
also missed in optical surveys. Very deep large field surveys in the
$K$ band ($K\sim 24$) are needed to provide an unbiased sample of high
$z$ galaxies.

\section {THE DWARF POPULATION AT INTERMEDIATE $z$}

\subsection {The luminosity function}

To compare in more detail our results with other surveys we have
computed the average volume density of galaxies with $19.5<R'<25$
derived in the redshift interval $0.5<z<0.75$ where we have
sufficient statistics and where the redshift estimate is less
uncertain due to the presence of the 4000 {\AA} break in the $V-R$
color. The luminosity function has been computed following Lilly et
al. (1995) using the $1/V_{max}$ formalism: $\phi(M,z)dM=\sum _k
1/V_{max}$.  Absolute rest frame blue AB magnitudes have been computed
directly from the luminosities of the best-fit model spectra. These
fluxes represent an interpolation between the observed $V'-R'$ colors
at $z\sim 0.6$.  Thus the spectral fit obtained for each galaxy
provides the appropriate k-correction for each spectral type in the
galaxy sample.  The resulting luminosity function is shown in Fig.~6a
in the magnitude interval $-20<M_{B_{AB}}<-17$.  For comparison we
have plotted the analytical fit (dotted line) obtained in the same
redshift interval by Lilly et al. in the CFRS sample at $I<22.5$.
Comparing our data with the CFRS data it appears that the luminosity
function is consistent with the average density of brighter surveys
like CFRS and Autofib/LDSS (Ellis et al. 1996) for $M_{B_{AB}}\leq
-19.5$ while it shows a sharp steepening for magnitudes fainter than
$M_{B_{AB}}=-19$. A density of $\sim 2\times 10^{-2}$ Mpc$^{-3}$
mag$^{-1}$ at $M_{B_{AB}}=-17.5$ is reached, in general agreement with
that derived in the HDF in about the same redshift interval by Sawicki
et al. (1997).

Bringing together the CFRS and our luminosity functions results in a
global LF which is not well represented by a Shechter shape since a
steepening over the Schechter faint-end slope derived from the CFRS
sample is apparent at $M_B>-19$ in our data.  This steepening is
reminiscent of that found in the local luminosity function by Zucca et
al. (1997) where a similar volume density of the order of $2\times
10^{-2}$ Mpc$^{-3}$ mag$^{-1}$ is obtained at magnitudes $\sim
M_{B_{AB}}=-15$ (see Fig. 6b). We note that, with a dimming of 2 magnitudes,
our LF is in good agreement with the local LF for $M_{B_{AB}} \geq
-17$ within the rather large errors. Thus a strong luminosity
evolution is suggested for the faint end of the luminosity function in
the redshift range $z=0-0.8$.  Since it is known from brighter surveys
that the luminosity evolution of $L^*$ galaxies is small (cf. Ellis
97), a cosmological scenario where a different population of dwarf
galaxies is subject to a strong luminosity evolution (Cowie et
al. 1991, Babul \& Rees 1992, Phillipps \& Driver 1995) appears
consistent with the LF evolution derived from this photometric
sample. Although more complex scenarios involving number density
evolution can not be excluded by the present data, larger and deeper
photometric surveys are needed to constrain the evolution of the faint
end of the galaxy LF.

\subsection {The physical properties of galaxies at intermediate $z$}

When the photometric estimate of the galaxy redshift is accurate
(i.e. $\sigma _z\leq 0.1$) the best-fit model predicts intrinsic
luminosities and related physical parameters like in particular age,
star-formation rate and mass for each galaxy in the sample.  The
estimates of the luminous stellar mass are not unique because depend
on crucial galaxy properties like the IMF adopted in the models. The
ages derived by different authors differ by as much as 30\%
(e.g. Charlot 1996, Charlot, Worthey \& Bressan 1996). In the
framework of the models adopted here we have found that photometric
uncertainties $\delta m\sim 0.1$ in the broad bands produce, at the
best-fit redshift, uncertainties by a factor 2 in the previously
mentioned physical parameters. Taking these uncertainties in mind we
concentrate in identifying statistical trends suggested by the
observations rather than in determining the true values of the
physical parameters for each galaxy in the sample.

Fig.~7 shows the distribution of the galaxy ages in the redshift
interval $z=0.4-0.8$. The distribution has a median age of $10^9$
yr. It is to note that flat spectrum, star-forming galaxies with
$B'-I'\leq 1.4$ as defined in color surveys (Cowie et al. 1996), show
a narrow distribution peaked at $2\times 10^8$ yr testifying the
presence of a population of very young star-forming galaxies.

Only a small fraction ($\sim 20$\%) of galaxies has been recognized
with a signature in their broad band spectra of a recent episode of star
formation superimposed to an older component (i.e. with a Scalo
parameter $b>1$). Most of these galaxies have $B'-I'> 1.4$. Although
they show some current star-formation activity, most of their luminous
stellar mass belongs to an older episode of star formation.

A second important parameter derived from our fitting technique is the
mass in stars present in the galaxies. This is tied to the absolute
magnitude in the K band, $M_K$, and it is relatively insensitive
(within the accuracy previously discussed in this Section) to the
accurate estimates of the galaxy age and metallicity.  The stellar
mass distribution for the galaxies in the redshift interval
$z=0.4-0.8$ where the LF has been computed, is shown in Fig.~8. The
distribution of the mass has a median value $\sim 5\times 10^8$
M$_{\odot}$. However the sample of the blue galaxies with $B'-I'\leq
1.4$ has a narrower distribution with a mean at $\sim 10^8$
M$_{\odot}$ and with no galaxies having $M>2\times 10^9$
M$_{\odot}$. This implies that the bulk of blue galaxies with $B'\sim
25$ has a very small luminous stellar mass.

In summary, considering the redshift distribution in Fig.~4 it appears
that the majority of galaxies at $z<1$ has a mass $M<2\times 10^{9}$
M$_{\odot}$ and an average age $\sim 10^9$ yr.

These results suggest that any pure luminosity model (PLE) based on a
single formation redshift (usually confined $z>2$) adopted for all the
faint field galaxies is inconsistent with the short age and small
mass determination of the blue redshift galaxies observed in the
$z=0.4-0.8$ redshift interval. Indeed the median formation redshift
derived for these star forming galaxies is $z_{form}\simeq 0.9$.

\section {THE STAR FORMATION HISTORY OF\\ THE FIELD GALAXIES}

The observed luminosity density at the rest frame $\lambda \sim 2800$
has been computed in each redshift bin as $\sum _i L_i/V_{max,i}$ with
$V_{max,i}$ for each galaxy $i$ defined as for the evaluation of the
luminosity function.  The cosmological trend in Fig.~9 and Tab.~3 shows an
increase up to $\phi _{2800} =2\times 10^{19}$ W Hz$^{-1}$ Mpc$^{-3}$
at $z=0.8-1$ and a gradual decline towards a value $\sim 10^{19}$ W
Hz$^{-1}$ Mpc$^{-3}$ at $z\sim 3-4$.

We can also estimate the cosmological metal production rate directly
from the observed luminosity density assuming dust free emission and
solar metallicity, in analogy with previous similar estimates
(e.g. Cowie et al. 1996; Madau et al. 1996, Connolly et al. 1997). The
values are reported on the same Fig.~9.

It is to note that the values shown in Fig~9 represent the actual
values measured in our samples. We have applied no correction to
account for galaxies fainter than our magnitude limit in contrast with
the values computed by Lilly et al. (1995), Madau et al. (1996),
Sawicki et al. (1997), Connolly et al. (1997) where a correction for
incompleteness by a factor 2.5 at $z\sim 1$ as derived from an
extrapolation of the luminosity function has been applied to the CFRS
data. In fact our values at $z=0.4-0.6$ are in very good agreement
with the observed values by Lilly et al. (1995) for galaxies with
$I_{AB}\leq 22.5$. Our observations indicate that the contribution by
fainter galaxies with $I_{AB}=22.5-25$ is small in this redshift
interval. The maximum value $\log \phi=19.25$ at $z=0.7-1$ is only
0.15 higher in our 2.5 mag fainter sample and in good agreement with
the value $\log \phi=19.3$ derived by Connolly et al. (1997) from the
deeper HDF sample. Thus small corrections are expected in the $0.3\leq
z\leq 1$ interval with respect to the values plotted in Fig.~9. At $z=1-2$
a comparison with the observed values derived from the HDF sample
($\log \phi \sim 19.5-19.4$; Connolly et al. 1997) suggests that our
brighter sample provides a luminosity density a factor 2 lower (Fig.~9).

We have also included in Fig.~9 the recent local ($z\simeq 0.15$) data
by Treyer et al. (1997) which update to a higher value the local UV
emissivity. The local UV emissivity is provided by an UV galaxy
luminosity function extended towards luminosities as low as the ones
in the $z\sim 0.5-0.8$ blue LF. Adding this new local value it appears
that the cosmological luminosity density changes by a factor $\sim
2.5$ in the overall redshift interval $z=0.1-4$. This factor can
increase to $\sim 3.5$ considering deeper results from the HDF. Thus
any evidence of a marked maximum in the luminosity density at $z\sim
1-1.5$ appears blurred especially if we consider that at $z>2$ we are
sampling only bright galaxies with $L_B\geq L^*$ and corrections for
the contribution by fainter galaxies become appreciable even in the
HDF sample.  A conservative estimate at $z=3$ gives $\log \phi \sim
19.4$ (Madau 1997). A decline appears at $z>3.5$ but the statistical
uncertainties at these very high redshifts are large.

At this point, rather than converting the observed UV luminosity
density in a star-formation rate by means of a fixed conversion
factor, we have exploited the information derived from the best-fit,
spectral models to compute the SFR in individual galaxies from their
intrinsic dereddened colors. 

The distribution of the star-formation rate per unit comoving volume
as a function of redshift is shown in Fig.~10.  The average mass
density formation rate increases from the local value $\sim 0.025$
M$_{\odot}$ yr$^{-1}$ Mpc$^{-3}$ (for a Miller-Scalo IMF and
$E_{B-V}=0.1$) up to $\sim 0.05$ M$_{\odot}$ yr$^{-1}$ Mpc$^{-3}$ in
the redshift interval $z=0.75-1$. At redshift $z\sim 3$ there is a
decline toward values $\sim 0.03$ M$_{\odot}$ yr$^{-1}$ Mpc$^{-3}$, 2
times lower than at $z=0.75-1$ but comparable to the local value. At
$z>3.5$ the SFR decreases progressively. It is to note that the galaxy
spectral models provide evidence for appreciable amount of dust
reddening, increasing the SFR distribution by a factor 2.5 in the
redshift interval $z=0.6-1$ (corresponding to a typical $E_{B-V}\sim
0.1$ using the Calzetti attenuation law) respect to the value derived
from the luminosity density evolution assuming no dust attenuation.
On the other hand, only a relatively small fraction $\sim 25$\% of the
galaxies in the field shows a color behaviour indicative of very low
metallicity, with $Z<0.1 Z_{\odot}$. These very blue galaxies provide
an appreciable UV luminosity density but a corresponding SFR 1.4
smaller than expected from stars with $Z\simeq Z_{\odot}$. We note
however that the fraction of galaxies showing very low metallicity
depends on the assumed IMF and on the accurate photometry available
from the B to the K band. This value should only be taken as
indicative of the presence of a significant fraction of
low-metallicity galaxies.

The same considerations done for the luminosity density evolution
apply to the cosmological evolution of the average SFR.  Any change in
the SFR in our sample is confined within a factor $\sim 2.5$ in the
redshift interval $z=0.1-3.5$.

Finally, the cosmological evolution of the stellar mass per Mpc$^3$ in
our sample increases for decreasing redshifts reaching a value $\sim
10^8$ M$_{\odot}$ Mpc$^{-3}$ at $z\sim 0.5$ (Fig.~11). This behaviour
follows the cosmic evolution of the SFR per comoving volume.
Considering the small age resulting for the galaxies with $z=3-4$, the
stellar mass density at very high redshifts is only 10--20\% of that
present at $z\sim 0.5$. On the other hand, the mass contribution at
these very high redshifts predicted by models based on hierarchical
clustering and merging is $\leq 7$\% at $z\geq 3$ for the Miller-Scalo
IMF (Baugh, Cole \& Frenk 1996), a value smaller although not far from
that derived from the observation.  However, since incompleteness in
the cosmological SFR (and correspondent stellar mass) at $z\sim 3-4$
by a factor $\geq 1.4$ (Madau 1997) could be present, an increasing
discrepancy with the values predicted by these models should be
expected.

\section{CONCLUSIONS}

Photometric redshifts have been obtained from the $R'\leq 25$ galaxies
selected in the field of the high redshift ($z=4.7$) quasar BR
1202-0725. The wide spectral coverage obtained from deep BVRIK
multicolor photometry has allowed an accurate redshift estimate for
each galaxy. This has been obtained comparing the observed colors with
those predicted by spectral synthesis models including UV absorption
by the IGM and dust reddening.  We have discussed the accuracy of the
method using a control sample of galaxies in the HDF with
spectroscopically confirmed redshifts. The main conclusions can be
summarized in the following items.

\begin{itemize}
\item
A comparison between spectroscopic and photometric redshifts for a
sample of galaxies selected in the Hubble Deep Field has shown that
spectral synthesis models (Bruzual \& Charlot 1997) including UV
absorption by the IGM and dust reddening can provide photometric
redshifts with comparable accuracy ($\sigma _z \leq 0.1$ at $z\leq 1.5$)
as obtained by means of the observed spectral templates of local
galaxies. The present approach has the advantage to exploit the
information on the spectral and physical properties derived from the
GISSEL models for each galaxy in the sample. An IMF shape steeper than
a Salpeter law provides unbiased redshift estimates and the following
statistical results are given for a Miller-Scalo IMF.
\item
The redshift distribution of the $R'\leq 25$ galaxies is peaked at
$z=0.6$ with 16\% of the sample at $z>1.5$.  The derived surface
density of the $z\sim 2.8-3.5$ galaxies having $\langle
M_{B_{AB}}\rangle=-21$ in our field is 1.6 arcmin$^{-2}$ in agreement
with that derived in the HDF (1.5 arcmin$^{-2}$) at about the same
magnitude. This corresponds to a comoving volume density of $10^{-3}$
Mpc$^{-3}$ similar to the local density of galaxies with the same
luminosities.  The derived surface density at $3.5<z\leq 4.5$ in our
field is lower, 1 arcmin$^{-2}$.
\item
The estimated luminosity function at $z\sim 0.6$ shows a strong
steepening for $M_{B_{AB}}>-19$ with respect to the extrapolation
derived from brighter redshift surveys. A comoving volume density of
$2\times 10^{-2}$ Mpc$^{-3}$ at $M_{B_{AB}}=-17.5$ is
obtained. Comparing with the local luminosity function, a luminosity
evolution by about 2 magnitudes is suggested for galaxies with
$M_{B_{AB}}>-19$.
\item
The bulk of the intermediate redshift population mostly consists of
very young star-forming galaxies with a median age $\leq 10^9$ yr and
a small stellar mass $M\sim 5\times 10^{8} $ M$_{\odot}$.  In
particular, the blue fraction with $B'-I'<1.4$ shows a median age of
$2\times 10^8$ yr and stellar mass $M\sim 2\times 10^{8}$
M$_{\odot}$. Any pure luminosity model (PLE) based on a single
formation redshift (usually confined to $z>2$) adopted for all the faint
field galaxies appears inconsistent with these small age and mass
evaluations.
\item
The observed 2800 {\AA} luminosity density and the associated star
formation rate in our sample show an increase to $\phi \sim 2\times
10^{19}$ W Hz$^{-1}$ Mpc$^{-3}$ (or $SFR\sim 5\times 10^{-2}$
M$_{\odot}$ yr$^{-1}$ Mpc$^{-3}$) at $z\sim 0.8$, i.e. only by a
factor 2.5 larger than the local value. At $z>1$ the UV luminosity
density and the corresponding SFR decrease to values comparable to the
local one. Thus evidence of a marked maximum in the luminosity density
and SFR at $z\sim 1$ appears blurred especially if we consider that an
significant corrections for fainter undetected galaxies are expected at
$z>1$. A comparison between the average cosmological luminosity
density and the corresponding star formation rate at $z=0.4-1$ implies
an average $E_{B-V}\simeq 0.1$, adopting the Calzetti (1997)
attenuation law and a Miller-Scalo IMF. Finally, the derived
cosmological mass in luminous stars per comoving volume at $z\sim 3-4$
is $\sim 20-10$\% of that formed at $z=0.5$, a value larger although
not far from that predicted by the standard hierarchical clustering
scenario.
\end{itemize}

\acknowledgments We thank S. Charlot for providing the most recent
version of the GISSEL library and for comments on an early version of
the paper, S. Arnouts for help in computing magnitudes for the HDF
galaxies, E. Zucca for providing data in advance of publication and
A. Renzini for useful discussions. This work was partially supported
by the ASI contract 95-RS-38 and by the Formation and Evolution of
Galaxies network set up by the European Commission under contract ERB
FMRX-CT96-086 of its TMR programme.

\bigskip
\bigskip
\centerline{\bf FIGURE CAPTIONS}

\noindent
Fig.~1 Colors as a function of redshift for a galaxy forming stars at
a constant rate from $z=5$. Circles, triangles, squares and exagons are
for $I-K$, $B-I$, $V-I$, $R-I$ AB colors, respectively. (Upper panel)
Filled symbols are for a galaxy with metallicity $Z=0.2 Z_{\odot}$,
open symbols for a galaxy with $Z=Z_{\odot}$. A Scalo IMF has been
adopted. (Lower panel) Filled symbols are for a galaxy with the
Salpeter IMF, open symbols for the Scalo IMF.

\noindent
Fig.~2 Photometric redshift estimates versus spectroscopic redshifts
for a subsample of galaxies in the Hubble Deep Field with $K$ band
photometry.  Three different IMF have been adopted as shown in the
Figure.

\noindent
Fig.~3 The spectrum of a galaxy forming stars at a constant rate at
$z=3.25$ or $z=4.25$.  The average attenuation due to the IGM
absorption is included. The Johnson $BV$, Gunn $r$ and
Kron--Cousin $I$ filters are shown as broken lines.

\noindent
Fig.~4 The photometric redshift distribution in our sample. Thick
histogram shows the $z$ distribution of galaxies with masses greater
than $2\times 10^{9}$ M$_{\odot}$.

\noindent
Fig.~5 Examples of broad band energy distributions of high
and intermediate redshift galaxies.  The best--fitting
spectrum is also shown.

\noindent
Fig.~6 a) The luminosity function of our sample in the redshift
interval $0.45\leq z\leq 0.75$ and in the magnitude interval
$-20<M_{B_{AB}}<-17$.  For comparison the analytical fit (dotted line)
obtained in the same redshift interval by Lilly et al. (1996) at
$I<22.5$ is also shown. b) The same luminosity function shifted by 2
magnitude lower. The local luminosity function by Zucca et al. (1997)
is also shown as empty circles. Their magnitudes have been scaled to
$H_o=50$ km s$^{-1}$ Mpc$^{-3}$ and to the AB system ($B_{AB}=B-0.16$).

\noindent
Fig.~7 The distribution of galaxy ages in the redshift interval
$z=0.4-0.8$. Thick histogram shows the distribution of the blue, star
forming galaxies with $B'-I'\leq 1.4$.

\noindent
Fig.~8 The mass distribution in the redshift interval $z=0.4-0.8$.
Thick histogram shows the distribution of the blue galaxies with
$B'-I'<1.4$.

\noindent
Fig.~9 The luminosity density distribution derived from our sample at
the rest frame wavelength of 2800 {\AA}. The local value (filled
circle) has been taken from Treyer et al. (1997). The directly
observed values from the CFRS sample by Lilly et al. (1996) and from
the HDF sample by Connolly et al. (1997) are shown as crosses and
empty circles, respectively.  The values derived from the $z>2$ HDF
sample (at $\lambda =1500$ {\AA}) by Madau et al. (1996) are also
shown as empty squares. All these values have been converted in metal
production rates (MPR) adopting the conversion factor as in Connolly
et al. (1997).

\noindent
Fig.~10 The distribution of the star-formation rate per unit comoving
volume as a function of redshift for a Miller-Scalo IMF.  The cross at
$z=0$ indicates the local value derived by Gallego et al. (1995).

\noindent
Fig.~11 The cosmological evolution of the total formed mass in stars
per unit comoving volume.

\end{document}